\documentclass[twocolumn,prl,aps,amsfonts,amssymb,amsmath,showpacs]{revtex4}
\usepackage{graphicx} 
\begin{document}

\title{Large variation in the boundary-condition slippage for a rarefied gas flowing between two surfaces}

\author{J. Laurent, A. Drezet, H. Sellier, J. Chevrier and S. Huant}

\affiliation{Institut N\'eel, CNRS and Universit\'e Joseph Fourier, BP 166, F-38042 Grenoble Cedex 9, France}

\date{\today}

\begin{abstract}
We study the slippage of a gas along mobile rigid walls in the sphere-plane confined geometry and
find that it varies considerably with pressure. The classical no-slip boundary condition valid at
ambient pressure changes continuously to an almost perfect slip condition in a primary vacuum. Our
study emphasizes the key role played by the mean free-path of the gas molecules on the interaction
between a confined fluid and solid surfaces and further demonstrates that the macroscopic
hydrodynamics approach can be used with confidence even in a primary vacuum environment where it is
intuitively expected to fail.
\end{abstract}

\pacs{47.15.Rq, 47.10.ad, 47.61.Fg, 07.79.Lh}


\maketitle


It is traditionally assumed \cite{Batchelor} that in a fluid flowing along a solid surface,
molecules nearest to the surface are globally stopped due to friction and collisions. This so
called no-slip boundary condition has been very successful in modeling macroscopic experiments and
it indeed forms one of the fundamental axiom of classical hydrodynamics. However, it has recently
been recognized that this standard condition is often not valid at submicro- and nanoscales
\cite{Tabeling,Lauga}. Furthermore, the hydrodynamic behavior close to a solid surface changes
drastically with interfacial phenomena like roughness or surface chemistry and the exact physics
underlying these variations is not well understood \cite{Tabeling,Lauga,Bocquet,MicroNanoFluid}.
Beyond its fundamental interest, elucidating these boundary conditions becomes a key issue for
micro- and nano-electromechanical systems (MEMS-NEMS) such as sensors and actuators working in
fluidic environments (liquid or gas). Although important results have been reported in a liquid
environment (e.g. flow through nano- and microchannels or nano-tribology
\cite{Tabeling,Lauga,Neto,Vinogradova2}) there are only a few indications to the significance of
these phenomena in gases \cite{Maali,Verbridge,Siria,Honig,Drezet}.\\
\indent In this letter, we analyze a simple model apparatus able to continuously tune in the
sphere-plane confined geometry the slippage boundary conditions at the solid-gas interface. By
decreasing the surrounding pressure of a sphere facing a rigid wall in a gas (air or Helium), we
find that these boundary conditions continuously evolve from a viscous regime supporting no slip to
a ballistic regime with perfect slip. We interpret our results in terms of a giant modification of
the gas slippage at the interfaces. Therefore, our experiments appear to reconcile in a single
setup boundary conditions that look conflicting at first sight. \\
%
\begin{figure}
\begin{center}
\includegraphics[width=\columnwidth]{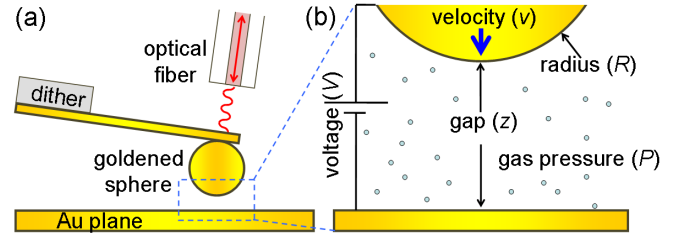}
\caption{(a) Scheme of the experimental setup. (b) Zoom on the interaction zone between the facing
surfaces defining the relevant physical parameters.} \label{figure1}
\end{center}
\end{figure}
%
\indent The experimental setup (Fig.~\ref{figure1}) is a homemade atomic force microscope (AFM)
working at 300\,K under controlled atmosphere. An optical-fiber based interferometric detection of
the cantilever \cite{Jourdan,Siria} provides the required sensitivity to measure the impact of gas
confinement on the viscous damping of the probe. The latter is a 460\,$\mu$m long, 50\,$\mu$m wide,
and 2\,$\mu$m thick, silicon AFM microlever with a $R=20\,\mu$m radius polystyrene sphere glued at
its extremity. In order to control the electrostatic interaction (see below), the whole probe (i.e.
cantilever and sphere) is coated with a 200\,nm thick gold layer, and the AFM chip is glued with
silver paint on a holder attached to the microscope frame. The probe spring constant $k=0.45$\,N/m
and the resonance frequency $f_0=9420$\,Hz have been measured using the Brownian motion of the
thermally actuated lever at 300\,K \cite{Siria,Drezet}. The planar surface facing the sphere is a
silicon substrate coated with a 200\,nm thick gold layer and mounted on a high-precision
positioning system to adjust the cavity gap $z$ between the sphere and the sample. An inertial
motor makes sub-micron steps over a large 7\,mm displacement range, whereas a piezo-scanner
corrected for hysteresis distortions ensures a fine vertical positioning over a 1.5\,$\mu$m range.\\
\indent For our measurements, it is important to conceive a method for measuring the absolute gap
$z$. This is achieved by applying a voltage bias $V$ to the probe with respect to the facing flat
surface as commonly done in Casimir force measurements where contact between facing surfaces must
be avoided \cite{Jourdan,deMan}. The cantilever is mechanically actuated with a dither at its
resonant frequency using a phase-locked loop device and the resonant frequency shift $\Delta
f=\alpha\;(V-V_c)^2$ is recorded as a function of $V$ to extract the coefficient $\alpha$ that
bears the desired distance information ($V_c$ is the difference in materials work functions). For
$z\ll R$ in the sphere-plane geometry, we use the relation $\alpha \approx f_0\pi\epsilon_0 R /
(2kz^2)$ to find the absolute distance $z$. In fact, to reach a better precision of 2\% on $z$, we
record $\alpha$ during a precise 1.5\,$\mu$m scan around the mean position and fit the results
against the above formula.\\
%
\begin{figure}
\begin{center}
\includegraphics[width=\columnwidth,clip,trim=10 10 10 10]{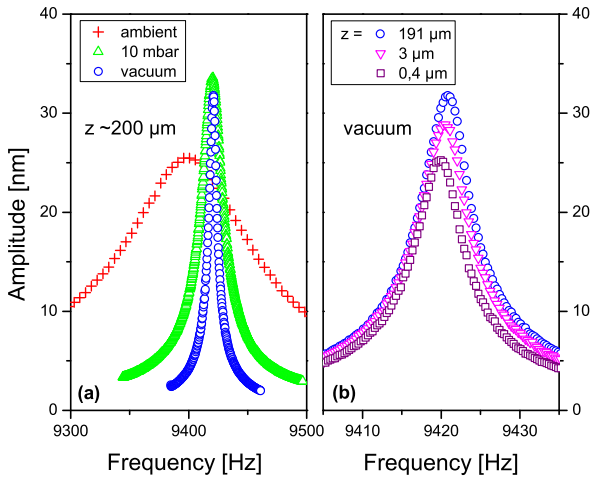}
\caption{(a) Resonance spectra of the cantilever-sphere probe at three different pressures: ambient
(1\,bar), helium gas (10\,mbar), and vacuum ($4\times 10^{-2}$\,mbar), when the fluid is not
confined (the sphere to surface distance is $z\approx 200\,\mu$m). Note that the mechanical
excitation has been reduced at low pressure to keep a similar amplitude at resonance. (b) Resonance
spectra of the probe in vacuum for three different gaps $z$. The resonance width, i.e. damping
factor, increases with decreasing separation.} \label{figure2}
\end{center}
\end{figure}
%
\indent We have measured the vibration amplitude $A$ as function of the excitation frequency $f$
for three gas pressures $P$ and several distances $z$. Examples of resonance curves are shown in
Fig.~\ref{figure2}(a) at large distance and for different pressures and in Fig.~\ref{figure2}(b)
for different distances in a primary vacuum. By fitting each curve with the Lorentzian response of
an harmonic oscillator:
\begin{eqnarray}
 A[f] = A_0\frac{\gamma f_0}{\sqrt{(f_0^2-f^2)^2+\gamma^2 f^2}}
 \label{eqn:response}
\end{eqnarray}
we obtain the dissipation constant $\Gamma=k\gamma/(2\pi f_0^2)$ of the probe, with a fitting error
smaller than 1\%. The results are shown on the left graphs of Fig.~\ref{figure3} for three pressure
conditions: (a) air at atmospheric pressure (1000\,mbar), (b) helium at low
pressure (10\,mbar), and (c) air vacuum ($4\times 10^{-2}$\,mbar).\\
\indent When the sphere is far from the sample, the damping factor reaches a constant value
$\Gamma_0$, characterizing the dissipation of the \{cantilever + sphere\} oscillating system in the
fluid. This value is directly extracted from the data and reported in Table~\ref{table} (except in
vacuum where the damping factor is not saturated at the largest distance and $\Gamma_0$ was
obtained by fitting with Eq.~\ref{eqn:vino} as described below). As is also visible in
Fig.~\ref{figure2}(a), $\Gamma_0$ decreases with decreasing pressure $P$ in agreement with previous
works~\cite{Verbridge,Li,Svittelskiy}. In the viscous regime this is mainly due to the existence of
a boundary layer of thickness $\delta_B\propto 1/\sqrt{P}$ \cite{Batchelor,noteb} representing a
dissipation channel at finite frequency that adds to the intrinsic losses $\Gamma_{\textrm{int}}$
of the lever (i.e. losses in the limit $f_0\rightarrow 0$):
$\Gamma_0=\Gamma_{\textrm{int}}+O(\sqrt{P})$~\cite{Verbridge,Li,Svittelskiy}.\\
%
\begin{figure}
\begin{center}
\includegraphics[width=\columnwidth,clip,trim=10 10 10 10]{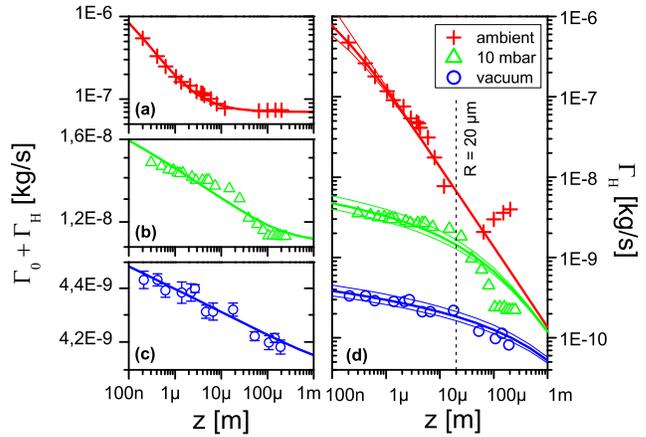}
\caption{(a-c) Evolution of the damping factor $\Gamma(z)=\Gamma_0+\Gamma_H(z)$ as a function of
the gap $z$ between the sphere and the plane at three pressures: (a) ambient air at 1\,bar, (b)
helium gas at 10\,mbar, and (c) air vacuum at $4.10^{-2}$\,mbar. The symbols represent the data (at
1\,bar and 10\,mbar, the experimental error bars are within the symbol size) and the continuous
lines are the fitting curves according to the theoretical model. (d) Evolution of $\Gamma_H(z)$
with the gap $z$ using the same color code as (a-c). The two curves (thin lines) surrounding each
fitting curve (thick lines) provide an error estimate on the slip length $b$ (from top to bottom:
$b=0$, 0.05, 0.1, 55, 70, 85, 1000, 1200, 1400\,$\mu$m).} \label{figure3}
\end{center}
\end{figure}
%
\indent When the gap decreases, the hydrodynamic force due to the gas confinement contributes a
$z$-dependent additional dissipation channel $\Gamma_H(z)$ such that
$\Gamma(z)=\Gamma_0+\Gamma_H(z)$. The central result of our work is shown in Fig.~\ref{figure3}(d)
where $\Gamma_H(z)$ is plotted for different pressures. It is clearly
seen that the dissipation observed at small distance is strongly reduced at low pressure. If in
agreement with usual statistical mechanics~\cite{Batchelor} we accept that the fluid viscosity
$\eta$ does not depend on pressure (a reasonable assumption in the viscous regime) we can conclude
that the boundary conditions at the solid-fluid interface should strongly change with pressure. In
other words, the so-called slip-length $b$, usually used to characterize the fluid flow at the
interface \cite{Lauga,Neto,Bocquet}, varies by a large amount. $b$ is related to the fluid velocity
gradient at the solid surface by $v|_\textrm{surface}=b\;\partial v/\partial z|_\textrm{surface}$
(where $v$ is the tangential fluid velocity) and can equivalently be interpreted as the fictitious
depth below the surface where the no-slip boundary conditions would be satisfied.\\
%
\begin{table}
\begin{center}\small
\begin{tabular}{|c|c|c|c|c|c|}
 \hline
 Gas & $P$ [mbar] & $\Gamma_0$ [kg/s] & $b$ [$\mu$m] & $\lambda_m$ [$\mu$m] & $p_d$ \\ \hline \hline
 Air & 1000 & $7.0\times 10^{-8}$ & 0.05 & 0.06 & 0.9 \\ \hline
 He & 10 & $1.12\times 10^{-8}$ & 70 & 10 & 0.17 \\ \hline
 Air & 0.04 & $4.1\times 10^{-9}$ & 1200 & 2500 & 1.2 \\ \hline
\end{tabular}
\caption{Measured slip length $b$ and asymptotic damping rate $\Gamma_0$ together with the
calculated mean free-path $\lambda_m$ and accommodation coefficient $p_d$ (see Eq.~\ref{eqn:slip})
at each pressure $P$.}\label{table}
\end{center}
\end{table}
%
\indent In the sphere-plane geometry with no-slip boundary conditions (i.e. $b\sim 0$), the
dissipation constant $\Gamma_H(z)$ is given in the limit $z\ll R$ by the Taylor formula
\cite{Batchelor,Neto}:
\begin{eqnarray}
 \Gamma_H(z)=\frac{6\pi\eta R^2}{z}
 \label{eqn:taylor}
\end{eqnarray}
where $\eta$ is the dynamic viscosity which does not change with pressure ($1.8\times
10^{-5}$\,kg.m$^{-1}$.s$^{-1}$ for air and $1.9\times 10^{-5}$\,kg.m$^{-1}$.s$^{-1}$ for Helium).
To take into account the gas slippage at the boundaries, we follow the reasoning of
Hocking~\cite{Hocking} and Vinogradova~\cite{Vinogradova1} and introduce a correction function
$f^*$ such that
\begin{eqnarray}
 \Gamma_H (z) = \frac{6 \pi \eta R^2}{z} f^*(\frac{z}{6b})
 \label{eqn:vino}
\end{eqnarray}
with $f^*(x) = 2x\left[\left(1+x\right) \ln{\left(1+\frac{1}{x}\right)} - 1\right]$. In this
formula (obtained for an incompressible fluid in the laminar regime~\cite{note}) we assume the same
values for $b$ on both surfaces since the two walls are coated with the same material. By fitting
the three sets of data in Fig.~\ref{figure3}(d) with Eq.~\ref{eqn:vino} in the range $z<20\,\mu$m,
we determine a slip length $b$ for each pressure. Note that $b$ is the only free parameter in the
fit, except for the data in vacuum where both $\Gamma_0$ and $b$ are simultaneous adjustable
parameters. $b$ is however not sensitive to the choice of $\Gamma_0$ since $b$ is determined
essentially by the absolute changes of $\Gamma(z)$. The results are presented in Table~\ref{table}
with an estimation of the error shown in Fig.~\ref{figure3}(d). The strong increase of $b$ at lower
pressure clearly shows that the friction of the confined fluid along the solid boundaries
considerably changes with pressure going from the usual no-slip condition in ambient air (i.e.
$\Gamma_H$ follows a $1/z$ law) to a quasi-perfect slip regime at low gas
pressure (i.e. $\Gamma_H \propto -\ln{(z/b)/b}$).\\
\indent In order to visualize the impact of a finite $b$ on the fluid dynamics close to the
surface, we compute the (radial) fluid velocity profile $v(Z)$ in the gap $z$ between the two
surfaces using the analytical theory of \cite{Hocking,Vinogradova1}. Due to the fluid
incompressibility and the limit $z\ll R$, the fluid is essentially ejected from the gap in the
radial direction in response to the vertical displacement of the sphere~\cite{notenote}. The
comparison between the three different values of $b$ obtained in the experiment (see
Table~\ref{table}) is shown in Fig.~\ref{figure4}. We clearly switch from the usual Poiseuille
parabolic velocity profile (i.e. $v\sim 0$ on the solid boundary) for $b\rightarrow 0$, to a
quasi-constant velocity profile in the gap for $b\rightarrow +\infty$. Therefore our experiment
reveals a continuous transition between these extreme regimes.\\
%
\begin{figure}
\begin{center}
\includegraphics[width=\columnwidth]{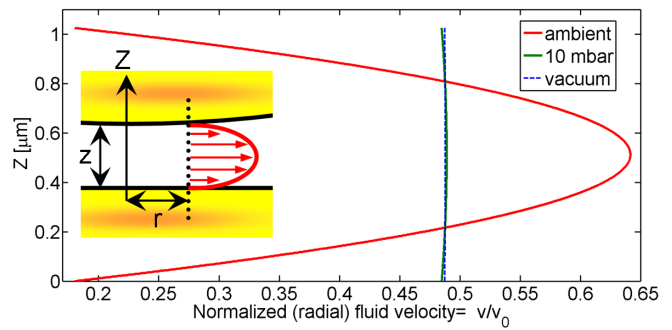}
\caption{The normalized radial fluid velocity profile $v(Z,r)/v_0$ ($v_0$ is the sphere velocity)
as function of the coordinate $Z$ in a $z=1\,\mu$m gap and for a radial coordinate $r=1\,\mu$m. The
profiles are calculated for the three $b$ values of Table~\ref{table}.} \label{figure4}
\end{center}
\end{figure}
%
\indent Now, from a microscopic point of view, the slippage coefficient is linked to the very
nature of the interaction between moving surfaces and air molecules. Following the historical
approach of Maxwell's kinetic theory of gases \cite{Maxwell}, two interaction channels can be
distinguished: a specular one where molecules collide elastically with the surface and a diffusive
one where molecules are reflected diffusively by the wall. The slip length $b$ in this statistical
model is given by the Maxwell formula \cite{Maxwell,Drezet,Sharipov,Fichman}:
\begin{eqnarray}
 b \simeq \frac{2}{3} \; \lambda_m \; \frac{2 - p_d}{p_d}
 \label{eqn:slip}
\end{eqnarray}
where $\lambda_m$ is the typical mean free-path of gas molecules and $p_d$ the tangential momentum
accommodation coefficient, i.e. the fraction of those particles hitting the surface with a
diffusive reflection.\\
\indent The main dependence of $b$ with pressure comes from the molecular mean free-path (see
Table~\ref{table}): $\lambda_m=\frac{k_BT}{\sqrt{2}\sigma P}$ where $T$ is the temperature, $k_B$
the Boltzmann's constant, and $\sigma$ the molecular cross-section. Two important asymptotic
regimes are clearly found in the experiment (see Fig.~\ref{figure4}). On the one hand, when the
mean free-path is extremely small compared to other macroscopic dimensions (i.e. at ambient air
pressure), the fluid particles interact strongly with themselves, and even more strongly with
surfaces located within a characteristic distance given by $\lambda_m$. This is a consequence of
multiple collisions, reflections, and adsorption processes \cite{Maxwell,Tabeling,Lauga,Bocquet}.
In this diffusive regime, the tangential velocity of the molecules decreases at short distance from
the surface, such that the slip-length $b$ tends to zero and the no-slip condition applies. On the
other hand, when the mean free-path is considerably larger than the gap, the molecules interact
mainly with surfaces and there is no momentum transfer among the molecules themselves. In this
ballistic or molecular flow regime, the velocity gradient vanishes at the surface. This results
in a giant slip length $b$.\\
\indent In Eq.~\ref{eqn:slip}, the accommodation coefficient $p_d$ (see Table~\ref{table}) depends
on the surface properties and gas density (i.e. on the probability of multiple collisions). Recent
analysis based on the fluctuation-dissipation theorem and the Green-Kubo relation emphasize the
importance of several microscopic parameters such as surface roughness and surface defects on the
molecular dynamics close to the surface \cite{Lauga,Bocquet,MicroNanoFluid,Neto}. In our experiment, the rms
roughness of the gold surfaces was found by AFM to be $\delta_r\simeq 3$\,nm for both the sphere
and sample. This certainly contributes to increase $p_d$ and reduce $b$ as compared to a perfect surface.\\
\indent Note that Eq.~\ref{eqn:vino} is an analytical solution of the Navier-Stokes equations whose
validity is well established when the fluid density is sufficiently large to provide a local
equilibrium within the fluid. However, in the present experiment in vacuum, the mean free-path
$\lambda_m$ is much larger than any other relevant length (e.g. the gap $z$) so that the system
enters in a molecular flow regime which should be discussed within the more general frame of
Boltzmann's kinetic equations \cite{Batchelor,Tabeling}. Therefore, the slip length $b$ deduced
from our data at low pressure should be considered as an extrapolation showing the limitation of
the usual hydrodynamic approach. Because the low-pressure data of Fig.~\ref{figure3} are well
reproduced in our analysis, our work stresses the remarkable robustness of the
Navier-Stokes Eq.~\ref{eqn:vino} even in this molecular regime. \\
\indent Finally, it is worth commenting on the differences between the present sphere-plane
experiment and recent results obtained in a plane-plane geometry \cite{Siria}. At atmospheric
pressure, we obtain here a small slip length in agreement with Refs.~\cite{Maali,deMan} whereas
Ref.~\cite{Siria} reports on a perfect slip in the same fluid (air). One of the difference is the
geometry, which is indeed known to be a critical parameter at micro- and nanoscales
\cite{Tabeling,Lauga,Li,Drezet,Siria2}. Another difference in Ref.~\cite{Siria} is the probe
velocity which resulted from the thermal motion of the cantilever with oscillation amplitude
$A\approx 0.05$\,nm, whereas the probe is here mechanically actuated i.e. $A\approx 30$\,nm (in the
same context compare Refs.\cite{Honig} and \cite{Maali}). In addition, the surface roughness was
only a few Angstroms, resulting in a smaller accommodation coefficient $p_d$ and a larger slip-length $b$.\\
\indent In conclusion, we have discovered that the slippage of a gas along  mobile rigid walls
varies considerably with pressure in the sphere-plane confined geometry. The classical no-slip
boundary condition valid at ambient pressure changes continuously to an almost perfect slip
condition in vacuum. This study emphasizes the key role played by the mean free-path $\lambda_m$ on
the interaction between a fluid and solid surfaces and demonstrates that the macroscopic hydrodynamics
approach can be used with confidence even in good vacuum conditions. We anticipate that our work will
have an impact on the MEMS and NEMS engineering and will motivate further fundamental studies of the
physics of gas slippage along solid and mobile surfaces.\\
\indent We are grateful to K.~Joulain and O.~Arcizet for helpful discussions and to A.~Mosset and
J.~-F.~Motte for valuable assistance during the experiments. This research was supported by the
PNANO 2006 program of the Agence Nationale de la Recherche under the project name ``MONACO''.




\begin{thebibliography}{99}
\bibitem{noteb}
The boundary layer $\delta_B=\sqrt{(\nu/f_0)}$, where $\nu$ is kinematic viscosity \cite{Batchelor}, is typically 20 $\mu$m at ambient pressure and reaches a few mm in vacuum.
\bibitem{note}
A typical value for the Reynolds number in our experiments is $R_e\sim f_0 A z/\nu\simeq 10^{-5}-10^{-6}$ at ambient pressure. At lower pressure this number is even reduced proportionally to pressure. This justifies our assumption of a laminar regime. Similarly, the gas incompressibility hypothesis is fully justified since the Mach number $M\sim f_0 A/c\simeq 10^{-6}\ll 1$ ($c$ is sound velocity) is very small.
\bibitem{notenote}
Here we have applied the Taylor formula (Eq.~\ref{eqn:taylor} and its
extension Eq.~\ref{eqn:vino}) to $z$ values exceeding the sphere radius $R$ although it is valid in the limit $z\ll R$ only,
so that our approach reveals general trends at large distances, rather than  offering an accurate quantitative analysis. Similarly, the small tilt angle ($\simeq 10^\circ$) of the vibrating cantilever with respect to the
plane surface (Fig.~\ref{figure1}) results in a small horizontal velocity (18\% of the main vertical
velocity) that we neglect, but this does not change the orders of magnitude discussed here.
\end{thebibliography}
\end{document}